\documentclass[onecolumn,amssymb,amsmath,amssymb,amsfonts,superscriptaddress,showpacs, article]{revtex4-1}

\usepackage{graphicx,graphics,pstricks,epsfig,wrapfig,amssymb,amsthm}
\usepackage{hyperref, longtable}

\begin{document}

\DeclareGraphicsExtensions{.pdf,.eps,.epsi,.jpg}

\title{}



\title{Antineutrino emission and gamma background characteristics from a thermal research reactor}

\author{V.M. Bui}
\affiliation{SUBATECH, CNRS/IN2P3, Universit\'e de Nantes, Ecole des Mines de Nantes, F-44307 Nantes, France}

\author{L. Giot \footnote{Corresponding author: giot@subatech.in2p3.fr}}
\affiliation{SUBATECH, CNRS/IN2P3, Universit\'e de Nantes, Ecole des Mines de Nantes, F-44307 Nantes, France}

\author{M. Fallot\footnote{Corresponding author: fallot@subatech.in2p3.fr}}
\affiliation{SUBATECH, CNRS/IN2P3, Universit\'e de Nantes, Ecole des Mines de Nantes, F-44307 Nantes, France}

\author{V. Communeau}
\affiliation{SUBATECH, CNRS/IN2P3, Universit\'e de Nantes, Ecole des Mines de Nantes, F-44307 Nantes, France}

\author{S. Cormon}
\affiliation{SUBATECH, CNRS/IN2P3, Universit\'e de Nantes, Ecole des Mines de Nantes, F-44307 Nantes, France}

\author{M. Estienne}
\affiliation{SUBATECH, CNRS/IN2P3, Universit\'e de Nantes, Ecole des Mines de Nantes, F-44307 Nantes, France}

\author{M. Lenoir}
\affiliation{SUBATECH, CNRS/IN2P3, Universit\'e de Nantes, Ecole des Mines de Nantes, F-44307 Nantes, France}

\author{N. Peuvrel}
\affiliation{SUBATECH, CNRS/IN2P3, Universit\'e de Nantes, Ecole des Mines de Nantes, F-44307 Nantes, France}

\author{T. Shiba}
\affiliation{SUBATECH, CNRS/IN2P3, Universit\'e de Nantes, Ecole des Mines de Nantes, F-44307 Nantes, France}

\author{A. S. Cucoanes}
\affiliation{SUBATECH, CNRS/IN2P3, Universit\'e de Nantes, Ecole des Mines de Nantes, F-44307 Nantes, France}

\author{M. Elnimr}
\affiliation{SUBATECH, CNRS/IN2P3, Universit\'e de Nantes, Ecole des Mines de Nantes, F-44307 Nantes, France}

\author{J. Martino}
\affiliation{SUBATECH, CNRS/IN2P3, Universit\'e de Nantes, Ecole des Mines de Nantes, F-44307 Nantes, France}

\author{A. Onillon}
\affiliation{SUBATECH, CNRS/IN2P3, Universit\'e de Nantes, Ecole des Mines de Nantes, F-44307 Nantes, France}

\author{A. Porta}
\affiliation{SUBATECH, CNRS/IN2P3, Universit\'e de Nantes, Ecole des Mines de Nantes, F-44307 Nantes, France}

\author{G. Pronost}
\affiliation{SUBATECH, CNRS/IN2P3, Universit\'e de Nantes, Ecole des Mines de Nantes, F-44307 Nantes, France}

\author{A. Remoto}
\affiliation{SUBATECH, CNRS/IN2P3, Universit\'e de Nantes, Ecole des Mines de Nantes, F-44307 Nantes, France}

\author{N. Thiolli\`ere}
\affiliation{SUBATECH, CNRS/IN2P3, Universit\'e de Nantes, Ecole des Mines de Nantes, F-44307 Nantes, France}

\author{F. Yermia}
\affiliation{SUBATECH, CNRS/IN2P3, Universit\'e de Nantes, Ecole des Mines de Nantes, F-44307 Nantes, France}

\author{A.-A. Zakari-Issoufou}
\affiliation{SUBATECH, CNRS/IN2P3, Universit\'e de Nantes, Ecole des Mines de Nantes, F-44307 Nantes, France}

\begin{abstract}
The detailed understanding of the antineutrino emission from research reactors is mandatory for any high sensitivity experiments either for fundamental or applied neutrino physics, 
as well as a good control of the gamma and neutron backgrounds induced by the reactor operation.
In this article, the antineutrino emission associated to a thermal research reactor: the OSIRIS reactor located in Saclay, France, is computed in a first part. 
The calculation is performed with the summation method, which sums all the contributions of the beta decay branches of the fission products, 
coupled for the first time with a complete core model of the OSIRIS reactor core. The MCNP Utility for Reactor Evolution code was used,
 allowing to take into account the contributions of all beta decayers in-core.
This calculation is representative of the isotopic contributions to the antineutrino flux which can be found at research reactors with a standard 19.75\% enrichment in $^{235}$U. 
In addition, the required off-equilibrium corrections to be applied to converted antineutrino energy spectra of uranium and plutonium isotopes are provided. 
In a second part, the gamma energy spectrum emitted at the core level is provided and could be used as an input in the simulation of any reactor antineutrino detector installed 
at such research facilities. Furthermore, a simulation of the core surrounded by the pool and the concrete shielding of 
the reactor has been developed in order to propagate the emitted gamma rays and neutrons from the core. 
The origin of these gamma rays and neutrons is discussed and the associated energy spectrum of the photons transported after the concrete walls is displayed.

\end{abstract}
\pacs{14.60.Pq, 23.40.-s, 24.10.Lx, 28.41.Ak, 28.41.-i, 28.50.Dr}
\maketitle


\section{Introduction}
%
This work is motivated by several physics goals, spanning fundamental and applied neutrino physics. 
Reactor antineutrinos are emitted in huge quantities by reactor cores, and this feature has been used by many particle physicists 
since the late fifties to study the neutrino properties\,\cite{refsNeutrinoProp}. 
In the seventies, was also born the idea that antineutrino detection could be used for reactor monitoring\,\cite{refNeutrinoMoni}. 
In a reactor core, the production of antineutrinos arises from the beta decays of fission products. 
The distribution of the latter nuclei are inherently linked to the structure of the nucleus undergoing the fission process and to the energy of the impinging neutron. 
Thus the resulting emitted antineutrinos reflect the composition of a reactor core, but also its thermal power. 
These latter features, combined with progresses in the R\&D of small antineutrino detectors and the development of dedicated reactor simulations, 
could be of interest for the International Atomic Energy Agency (IAEA) in its task of safeguarding nuclear reactors worldwide\,\cite{IAEAReport, PRLHuber}. 
In parallel, with the advance of the next generation of reactor antineutrino experiments, 
new calculations of the main uranium and plutonium isotope antineutrino energy spectra have been recently developed\,\cite{Mueller, Huber} and 
led to the so-called "reactor anomaly``\,\cite{Mention}. 
Several hypotheses have been raised to explain this anomaly, among which the existence of light sterile neutrinos\,\cite{WhitePaper}.
Since then, short baseline experimental projects have been born 
at research reactors in order to evidence the existence of such sterile neutrinos, in addition to the ones already under development 
for applications of neutrino physics\,\cite{RevueNonProlif},
 or other physics goals such as neutrino nucleus coherent scattering experiments \,\cite{NNCS}, or the measurement of the neutrino magnetic moment\,\cite{NeutrinoMagneticMoment}.

Originally designed to be a small detector dedicated to reactor monitoring with an optimized detection efficiency, the Nucifer experiment\,\cite{NuciferPaper} 
is one of those experiments that could as well bring additional constraints on the phase space associated to a sterile neutrino in the eV$^2$ range. 
This experiment has provided the framework of the results presented in this article: a simulation of the OSIRIS reactor core close to which Nucifer is installed, a prediction
 of the associated antineutrino emission, as well as of the gamma and neutron backgrounds generated by the reactor operation.
Most of worldwide research reactors use fuel enriched up to a maximum of 20\% in $^{235}$U, which is the upper limit to be qualified as low enriched uranium
 and is strongly supported by the International Atomic Energy Agency \cite{IAEA}.   
This enrichment guarantees that most of the fissions arises from $^{235}$U thermal fission, but there may be a small contribution to the fissions coming 
from the $^{239}$Pu produced during irradiation from 
the 80\% of $^{238}$U composing this kind of fuel.
 This contribution has to be quantified as the $^{239}$Pu antineutrino energy spectrum shape differs from the one from $^{235}$U, 
 which requires the modelling of the reactor core and of the subsequent antineutrino emission. 

The reference antineutrino spectra are still the ones based on the integral beta spectra precisely measured at the high flux reactor of the Institut Laue Langevin in the eighties
 \cite{SchreckU5-1,SchreckU5-2,SchreckU5Pu9,Schreckenbach}, and lately the integral beta spectrum produced through fast fission of $^{238}$U has been measured in Munich
  at the Garching reactor\,\cite{PRLHaag}.
Currently antineutrino spectra obtained through the conversion of these spectra\,\cite{Mueller,Huber} 
are the ones exhibiting the smallest systematic errors. 
However, it should be mentioned that these systematic errors are the object of numerous discussions, 
showing that there may be extra sources of systematics to be accounted for due to nuclear physics theoretical and experimental uncertainties, insufficiently considered up to now \cite{Hayes}.
 The recent observation of a distorsion of the antineutrino energy spectra\,\cite{Neutrino2014} measured by Double Chooz\,\cite{DC}, Reno\,\cite{Reno} 
 and Daya Bay\,\cite{DB} 
 above 4\,MeV with respect to converted spectra 
 reinforces the need for further investigations. Reactor neutrino experiments have up to now mostly used converted spectra as predictions to compare their measurements to. 
 As these spectra were measured after a short irradiation time of a target into a research reactor, an extra correction has to be applied to account for the "off-equilibrium effects", 
 due to the build-up of long lived fission products, and to neutron captures on fission products during the core cycle. 
An alternative to this methodology is to use the summation method\,\cite{Fallot, Mueller, PRL2015}, consisting in summing the contributions of all beta decayers to the antineutrino spectrum. 
Up to now, the published summation method spectra were computed to match the conditions of the individual beta energy spectra of the main uranium and plutonium isotopes, measured at
 ILL\,\cite{SchreckU5-1,SchreckU5-2,SchreckU5Pu9,Schreckenbach}  and Garching\,\cite{PRLHaag}, i.e. not under the physics conditions of a reactor core under operation.
In this article, we have coupled the summation method for the first time with a complete reactor model, allowing us predicting the low energy part of the antineutrino spectrum, 
below the Inverse Beta Decay process (IBD) detection threshold, and evidencing the important contribution of actinides to the antineutrino emission. 
The obtained antineutrino energy spectra span an energy range beyond the one of the measured ILL and Garching integral beta spectra, which is the energy range of the converted spectra, 
making of the summation method spectra the only available predictions in these energy regions (below 1.8\,MeV and beyond 7-8\,MeV). 
In addition, are provided in this paper the corresponding off-equilibrium corrections, deduced for the first time from the simulation of a complete reactor core,
 that can only be computed with the summation method.

Experimental conditions at short distance from research reactors are challenging, 
because the reactor itself produces huge gamma and neutron backgrounds that induce accidental and correlated backgrounds in an antineutrino target. 
The understanding of these backgrounds is of utmost importance and triggered the second part of the work presented here. 
We have performed the simulation of the gamma rays and neutrons generated by the OSIRIS research reactor and propagated them through the water 
and concrete shieldings which separate the core from the antineutrino detector. 
The background estimation at the core level presented in this article could be used for the optimization of the background rejection of other experimental setups, 
or to estimate the physics reach of a foreseen experiment in the phase space of the oscillation angle and squared mass difference. 


%
%
In summary, we have thus developed a model of the OSIRIS research reactor core in order to: give orders of magnitude values of fission rates in small thermal research
 reactors using fuel enriched at about 20\,\% in $^{235}$U ; provide the required data 
to compute the emitted antineutrino energy spectrum, including fine effects such as off-equilibrium corrections; and eventually give estimates 
of the reactor-generated neutron and gamma backgrounds and identify their production processes.
The paper is organized as follows. A first section describes the OSIRIS research reactor and its fuel elements.
A second section describes the simulation tools which were used to perform these studies and the associated validation tests. 
In a third section, are presented the obtained results for the antineutrino energy spectrum and flux. 
The last section is dedicated to the neutron and gamma background studies.  
In Appendix A, are provided the antineutrino energy spectrum computed with the complete OSIRIS simulation and the summation method for a reactor at equilibrium and
off-equilibrium effects.
In Appendix B, we provide a table containing the gamma energy spectrum emitted by the core, 
before its propagation through the specific shielding of the OSIRIS reactor, 
so that it can be used as an input of a Monte Carlo simulation of another experimental site.   
 
\section{The OSIRIS Research Reactor}

The OSIRIS research reactor is an experimental reactor with a thermal nominal power of 70 MW \cite{OSIRIS}.
It is a light water reactor with an open core (57x57x60\,cm$^3$) which main objective is to irradiate, with a high neutron flux, 
samples such as fuel elements and structure materials 
from power plants, or to produce radioisotopes for medicine and industry.
The reactor delivers a neutron flux of 10$^{18}$ neutrons.m$^{-2}$.s$^{-1}$ for about 200 days per year as an average, 
with cycles of 3 to 5 weeks and reactor stop periods of 10 days between cycles.

\begin{figure}[hbt!]
\includegraphics[scale=0.3]{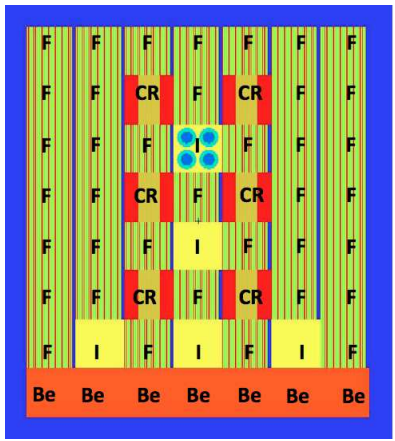}
\caption{\label{fig:Geo} View in the horizontal plane of the geometry of the simulated OSIRIS core vessel with MCNPX 2.5.e\,\cite{MCNPX}. 
The colours represent the different materials and universes defined in the core model. 
One can see the structure of the fuel elements (F), the Beryllium walls (Be), the control rods (CR) and the irradiaton locations (I) 
with one filled in with 4 Chouca ovens.}
\end{figure}

The core vessel with its rectangular cross-section is made of 4 cm thick Zircaloy and is surrounded by a 11~m deep, 7.5~m long and 6.5~m wide pool. 
The core vessel contains a rack of 56 cells used for: fuel elements, reflectors, control rods and experimental devices. Its geometry is displayed 
in Fig.~\ref{fig:Geo}.
The core is loaded with 38 standard fuel plate elements (F), 6 control command rods (CR), 5 locations for irradiation experiments (I).
Two safety rods ensure the reactor stop while 4 rods are compensative and operating rods.
An internal wall (7 cells) and an external wall to the vessel (5~cm thick) in Beryllium are located on the southern face of the core and act as reflectors.
Each of the five experimental positions in the core contains a water box which can house four tubes containing for each a device called a "Chouca oven" used for irradiation experiments.
The fuel elements are made with 22 standard fuel plates in a U$_{3}$Si$_{2}$Al alloy, which are enriched in $^{235}$U up to 19.75\% and conform to the IAEA requirements.
Each fuel plate is 0.51~mm thick and encapsulated in an aluminium cladding of 0.38~mm thick. The coolant channel between the plates is about 2.46~mm.
Two edge plates with boron are added to the fuel elements to comply with the regulatory safety margins and to control the reactivity available at the beginning 
of irradiation cycle.
The upper part of the control command rods is made of an absorbant material (Hafnium) and the lower part consists in 17 standard fuel plates, 
with a cooling channel of 2.79~mm between them. 
The Nucifer antineutrino detector is placed at 7~m from the core behind a 2~m thick concrete wall. 
The detailed description of the design of the detector can be found in\,\cite{NuciferPaper}.

\section{The MURE simulation}

In order to study either the core physics or the core-generated gamma 
and neutron backgrounds, the simulation of the reactor was made in several steps: the core alone (C), the core surrounded by the pool (CP) 
and the core surrounded by the pool 
plus a concrete wall (CPW) up to the casemate where the antineutrino detector is located.
The core simulation, cf. Fig.~\ref{fig:Geo}, has been performed with 
the MURE (MCNP Utility for Reactor Evolution) code coupled to the MCNPX 2.5.e Monte-Carlo code \cite{MURE,MCNPX}.
All the geometrical parameters were taken from public literature \cite{OSIRIS,Malouch}.
In the simulation, fuel elements are discretized into evolving cells and the evolution is temporally discretized. 
At each time step, an MCNP run is performed in order to compute neutron flux and energy dependence in each evolving cell. 
The neutron flux is then used, based on a Runge-Kutta 4$^{th}$ order method, to solve the Bateman equations which describe the fuel evolution. 
For each evolving cell, MCNP estimates a neutron flux normalized per neutron source. 
A scaling of the neutron flux in the cells is then performed to normalize it in order to reach the total thermal power value of the reactor given as an input,
 taking into account the energies released per fission taken from Kopeikin {\it et al.}\,\cite{Kopeikin}.
Considering the increase of the CPU time of the simulation due to a large spatial discretization of the geometry, assumptions were made on the discretization
 of the fuel elements in evolving cells.
No axial discretization was considered for all the reactor simulations performed during this work. Indeed, such axial dependence of the fission rates 
is very dependent on the operation of the reactor core with the control rods, which makes studies specific to the considered reactor 
site mandatory for each concerned neutrino experiment. This would be at variance with the objective of our paper - providing generic predictions 
and information about the antineutrino and background emission - which could be useful for a wider public than a specific experiment. 

Two core models were studied and compared.  
In the first model, all the fuel elements are treated as a unique evolving material cell, called a universe.
In the Monte-Carlo code MCNP used in the MURE code, the energy is treated continuously for the description of the associated interaction cross sections
 of the different considered particles.
The input nuclear cross sections used in this simulation were preferentially taken from the ENDFB/VI library and a thermal S($\alpha$,$\beta$) 
treatment was applied to the interaction of neutrons with water \cite{ENDFBVI}. 
A multigroup option of the MURE code was used to gain CPU time, consisting in discretizing the neutron energy distribution of the nuclear cross sections in  
1.79 10$^{5}$ groups. This option was tested in the frame of \cite{Jones, SimuDC} and showed results in agreement at the level of 0.2\% with the results obtained 
with a continuous energy treatment for the main fission rates of interest (uranium and plutonium isotopes) while gaining a factor of 30 in the CPU time
 of the calculation. 
In order to ensure a proper convergence of the neutron source in the geometry, a distribution of source neutrons is propagated in the geometry with 
an MCNP simulation using 50 000 source particles, 1 000 inactive cycles and 300 active cycles. 
The generated source is then used as an input in 
the subsequent simulations, with 20 000 source particles, 100 inactive cycles and 250 active cycles. 
In the second core model, confidential data, especially more detailed informations on the geometry\,\cite{RefCommOSiris} are used, and each 
of the 38 fuel assemblies is considered as an independent evolving cell,  i.e. 38 fuel universes are considered.

In order to validate the ability of these simulations to reproduce correctly the core physics, the neutron energy spectrum in an irradiation cell computed 
with MURE is compared to the results of\,\cite{Malouch}.
Material samples to be irradiated at OSIRIS are 
placed in experimental devices called "Chouca ovens" designed to reproduce the operational conditions of a power reactor.
These devices (32~mm in diameter) are rods made of an alloy of Zircaloy and stainless steel, surrounded by NaK and stainless steel layers.
They are inserted by groups of 4 in the water boxes. 
To study the impact of a converter device on the neutron flux in irradiation experiments, F. Malouch\,\cite{Malouch} computed the neutron energy spectrum 
in a Chouca oven by developing a 2D simulation of the OSIRIS core 
with the code APOLLO2 which solves explicitely the neutron transport Boltzmann equation on a spatial mesh and 
using the CEA93 cross-sections library consisting in 99 neutron energy groups \cite{Apollo2}.  
The CEA93 cross-sections library is processed from the JEF-2.2 evaluations \cite{Apollo2,JEF2}.
This model of the OSIRIS core has been validated with experimental measurements performed at the OSIRIS reactor. 
For this reason, we consider it as a reference with which we can compare our models.
\begin{figure*}[ht!]
\begin{center}s
\includegraphics[scale=0.6]{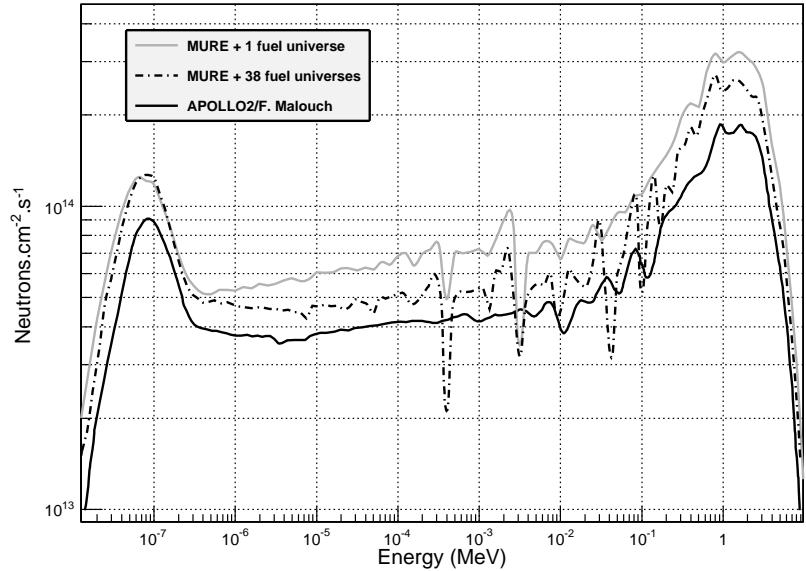}
\caption{\label{fig:ChoucaFlux} Neutron energy spectrum in a Chouca oven computed with APOLLO2 (solid black line extracted from \cite{Malouch}) and with the MURE code assuming 1 universe for all fuel elements (solid grey line) or 1 universe per fuel element (dashed black line).}
\end{center}
\end{figure*} 
%
%
In Fig.~\ref{fig:ChoucaFlux}, the neutron energy spectrum in a Chouca oven computed with APOLLO2 (solid black curve) is compared with the one calculated with MURE with the first core model (solid grey curve) 
on the 5$^{th}$ day of the power cycle assuming: a core loaded with fresh fuel,
a constant nominal power history and all the fuel assemblies grouped in a unique universe, cf Fig.~\ref{fig:Geo}, and with the second core model (dashed black curve) under the same conditions except that 38 fuel universes are considered.
The overall shape of the neutron energy spectrum is well reproduced by the MURE calculation.
The ratio betwen the neutron fluxes in the thermal and rapid energy ranges is 2.56 with the first MURE calculation, and 2.06 with the second one, compared with a ratio of 2.05 with APOLLO2. 

The normalization discrepancy arises from the unknown fuel loading map and the operation history associated to the APOLLO2 simulation. Especially no details
are available on the power history or on the different energies released per fission taken in the APOLLO2 calculation. 
Another origin of differences among the results of the APOLLO2 and MURE simulations arises also from the use of multigroup versus continuous in energy nuclear databases in respectively deterministic and Monte-Carlo codes,
as well as the use of more recent nuclear databases in the case of the MURE simulation with respect to the APOLLO2 one. This feature can be seen in Fig.~\ref{fig:ChoucaFlux} in the anti-resonant behaviour 
of the neutron energy spectrum between a few hundreds of eV up to a few hundreds of keV. 
Overall, the good agreement obtained on the shape of the neutron spectrum validates our core models.

\begin{figure*}[ht!]
\begin{center}
\includegraphics[scale=0.6]{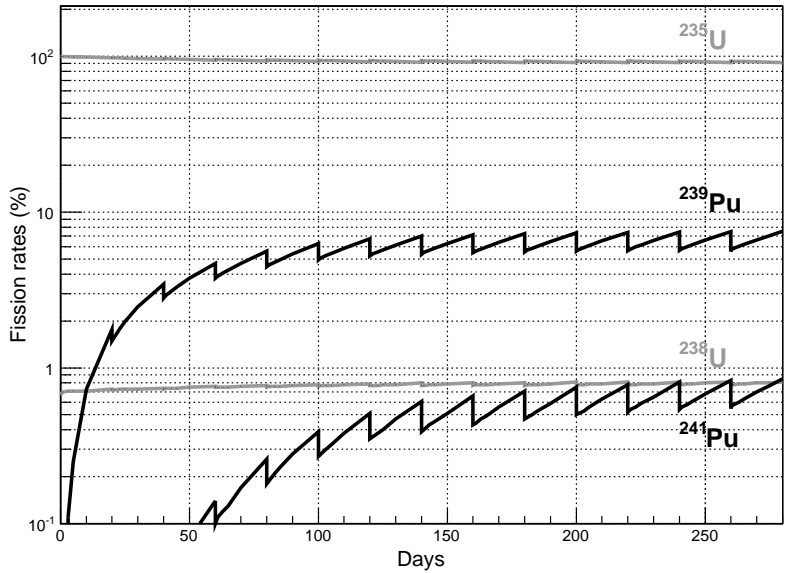}
\caption{\label{fig:FissionRates} Distribution of the fission rates in percent for the 4 main fissile nuclei: $^{235}$U, $^{238}$U, $^{239}$Pu and $^{241}$Pu taking into account a refueling scenario of 20 day period.}
\end{center}
\end{figure*}

As our objective is to simulate a typical research reactor antineutrino flux, a periodic refueling of the core was implemented.
The fuel evolution of the OSIRIS reactor was simulated over 200 days with a refueling of the core every 20 days by 
seven-th which corresponds to the length of a typical OSIRIS reactor cycle.
The CPU cost of such a detailed simulation imposes to simplify it.
To this purpose, we have grouped the assemblies in seven universes in order to take into account the refueling
of the core by one seven-th every twenty days. 
The refueling scenario is described on Table ~\ref{table:refueling}. The initials CR and I refer respectively to the location of 
the Control
Rods and Irradiation cells. The numbers (one to seven) refer to the refueling order of the fuel in the reactor core. 
At each refueling, the fuel labeled 1 is replaced by the one labeled 2 and so on until 6. 
The last fuel block (7) is replaced by fresh fuel. 
The cooling time of the fuel blocks between real reactor cycles is neglected in the calculation of the reactor evolution. 
Indeed this cooling time would affect mainly the fission rate of $^{241}$Pu which contribution to the fissions is very small, 
considering the enrichment of the fuel.
\begin{table}[!ht] 
  \centering
\begin{tabular}{|*{7}{p{0.5cm}|}}
\hline
\bf \centering 6 & \bf \centering 6 & \bf  \centering 5  & \bf  \centering 5 & \bf \centering 5  & \bf \centering 7 & \bf \centering 7  
 \tabularnewline
 \hline 
 \bf \centering 6 & \bf \centering 2 & \bf \centering {\tiny CR}  & \bf \centering 3  & \bf \centering {\tiny CR} & \bf \centering 2 & \bf \centering 7  
 \tabularnewline
 \hline 
 \bf \centering 4 & \bf \centering 3 & \bf \centering 1   & \bf \centering I  & \bf \centering 1 & \bf \centering 3 & \bf \centering 4  
 \tabularnewline
 \hline 
 \bf \centering 4 & \bf \centering 2 & \bf \centering {\tiny CR}  & \bf \centering 1  & \bf \centering {\tiny CR} & \bf \centering 2 & \bf \centering 4  
 \tabularnewline
 \hline 
 \bf \centering 4 & \bf \centering 3  & \bf \centering 1 & \bf \centering I  & \bf \centering 1 & \bf \centering 3 & \bf \centering 4 
 \tabularnewline
 \hline 
 \bf \centering 7 & \bf \centering 2  & \bf \centering {\tiny CR} & \bf \centering 3  & \bf \centering {\tiny CR} & \bf \centering 2 & \bf \centering 6   
 \tabularnewline
 \hline 
 \bf \centering 7 & \bf \centering I & \bf \centering 5 & \bf  \centering I & \bf \centering 5  & \bf \centering I & \bf \centering 6  
 \tabularnewline 
 \hline
 \end{tabular}
\caption{Refueling scenario chosen for the MURE calculation.}
\label{table:refueling}
\end{table}
The distribution of the fission rates for the 4 main fissile nuclei: ($^{235}$U, $^{238}$U, $^{239}$Pu and $^{241}$Pu) is presented in Fig.~\ref{fig:FissionRates} 
for an evolution of 280 days and 13 refuelings by seven-th. The fission rates are constrained by the power maintained at its nominal value along the evolution. 
By definition, the equilibrium of the reactor is reached when the fission rates show a periodic pattern repeated every 20 days. 
This is reached for the different contributions after 260 days, i.e the beginning of the third refueling cycle for the different fuel blocks. 
For a reactor with this fuel composition at equilibrium, the main contributor to the total amount of fissions in-core is $^{235}$U. 
For this isotope, the variation of the fission rate between the beginning and the end of the cycle is about 2-3\%.
Antineutrino emission thus mainly arises from the fission 
of $^{235}$U as it will be discussed in the next section.

%
%

\section{Associated Antineutrino Emission}

The antineutrino energy spectrum and flux are computed using the summation method\,\cite{Mueller, Fallot} using the MURE code coupled to nuclear databases containing 
beta decay properties of the fission products. The principle is to sum all the contributions of the beta decay branches of the fission products 
to obtain the total antineutrino spectrum, weighted by the fission product concentrations computed with MURE. 
In this article, the nuclear databases are the same as the ones presented in \cite{Fallot}, 
including the latest published Total Absorption Spectroscopy experimental results \cite{Algora} and \cite{PRL2015}. 
In the subsections below, we present antineutrino energy spectra computed in two different ways, using in both cases the summation method. 
In the first case, the complete core model is used to provide the amount in-core of all beta minus emitters, allowing for the first time to take into account all the contributions, i.e. not only the fission products but also the contribution of the actinides and heavy nuclei produced during core operation. 
The $\beta$/$\bar{\nu}$ energy spectrum emitted by the reactor core is thus broken-up into the sum of all $\beta$/$\bar{\nu}$ spectra of the beta minus emitters (labelled $be$) in-core, weighted by their activity $A_{be}$:
\begin{equation}
S(E,t)=\sum_{be=1}^{N_{be}}{A_{be}(t)\times S_{be}(E)}
\label{Sk}
\end{equation} 
Eventually, the $\beta$/$\bar{\nu}$ spectrum of one beta emitter is the sum over the b branches of all $\beta$ decay spectra (or associated $\bar{\nu}$ spectra), $S_{be}^{b}$ (in eq~\ref{Sfp}), of the parent nucleus to the daughter nucleus weighted by their respective branching ratios as
\begin{equation}
S_{be}(E)=\sum_{b=1}^{N_{b}}{BR_{be}^{b}\times S_{be}^{b}(Z_{be},A_{be},E_{0 be}^{b},E)}
\label{Sfp}
\end{equation}

In the second case, the complete core model is only used to provide the fission rates, which are then weighted with individual uranium and plutonium isotope spectra computed with the summation method, to obtain the total spectrum emitted by the OSIRIS reactor. 
The latter method is similar to the one used by reactor neutrino experiments i.e. weight individual antineutrino spectra from the main uranium and plutonium isotopes with their fission rates\,\cite{DC, DB, Reno} in the following way:
\begin{equation}
N_{\nu}(E,t)={\frac{P(t)}{\sum_{k}{\alpha_k(t)E_{k}}}}{\sum_k{\alpha_k(t)S_k(E)}}.
\label{eq_nu_flux}
\end{equation}
The first term accounting for the number of fissions occuring over the time $t$ is the ratio of the thermal power over the average energy released per fission of the four isotopes present in the fuel. $\alpha_k(t)$ stands for the percentage of fissions undergone by the isotope $k$ and is computable by reactor evolution codes. Its mean energy released per fission, $E_k$, is stored in nuclear databases. 
The second term of eq.~\ref{eq_nu_flux} represents the total $\bar{\nu}$ spectrum emitted by a reactor per fission. It is defined as the sum over the 4 isotopes ($^{235;238}$U and $^{239,241}$Pu) of the fraction of fissions undergone by the k$^{th}$ isotope times the $\bar{\nu}$ spectrum per fission of the same isotope $S_k(E)$.
In this paper, the $\beta$/$\bar{\nu}$ spectrum per fission of a fissible isotope $S_k(E)$ is calculated with the summation method, i.e. it is the sum of all fission product $\beta$/$\bar{\nu}$ spectra weighted by their activity $A_{fp}$, 
\begin{equation}
S_k(E)=\sum_{fp=1}^{N_{fp}}{A_{fp}\times S_{fp}(E)}
\label{Sk}
\end{equation} 
where each $S_{fp}$ is computed as in Eq.\,\ref{Sfp} above.

The individual spectra $S_k(E)$ are computed after an irradiation time corresponding to or close 
to the times for which converted spectra were obtained from ILL and Garching measurements i.e. 12\,hours for $^{235}$U and 36\,hours for the three other fissible nuclei, and without the contribution of neutron captures on the fission products, as in the ILL experimental conditions. 

The impact of the evolution of the concentration of antineutrino emitters in-core due to reactor physics and nuclide half-lives can be deduced from the discrepancy between the two calculations of the total antineutrino energy spectrum emitted by the reactor presented above. 
Using the summation method in both cases should allow to eliminate a large part of the systematic errors due to the calculation method as all correlated errors vanish in the relative discrepancy between both calculations.
In this way, the deduced effects are also directly applicable to correct the converted spectra used for computing the total antineutrino flux by the reactor neutrino experiments. 

In the next two subsections, are presented the obtained emitted antineutrino flux and energy spectrum and an example of flux and energy spectrum detected using the IBD process with a small liquid scintillator detector doped with Gd, placed at 7\,m from the OSIRIS reactor core with a detection efficiency of 50\%.

\subsection{Emitted Antineutrino Flux and Spectrum}
\begin{figure*}[hbt!]
\begin{center}
\includegraphics[scale=0.6]{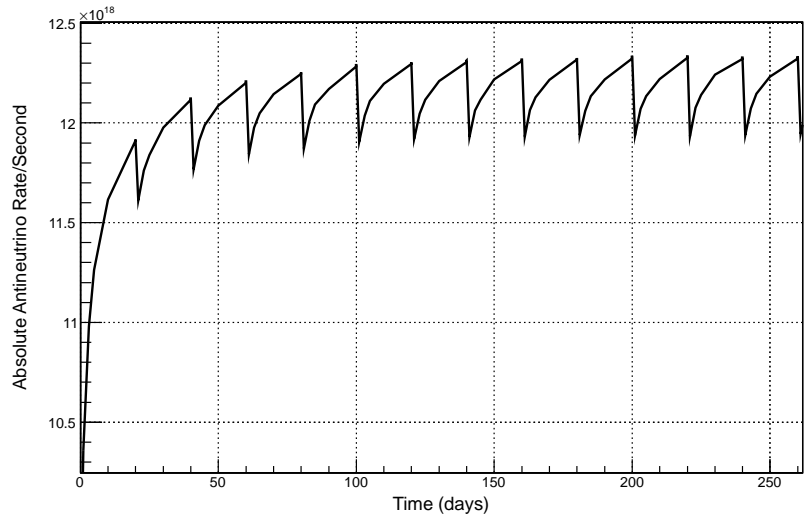}
\caption{\label{fig:AntinusEmis} 
 Antineutrino flux emitted by a research reactor with a $^{235}$U enriched fuel of 19.75\% computed with the summation method.}
\end{center}
\end{figure*} 
In Fig.\,\ref{fig:AntinusEmis}, the summation is made over the fission product concentrations computed with the OSIRIS core simulation described above, 
and includes the contributions of the beta decays of actinides which contribute to the energy spectrum at low energy, below the detection threshold 
of 1.8 MeV of the IBD process. 
It includes also off-equilibrium effects due to  neutron capture on fission products or to the evolution of the fission product amounts after core start-up. 
According to this simulation, the variation of the antineutrino flux emitted along the fuel cycle is of about 2-3\%.
 This variation is weak because of the important enrichment of the fuel in $^{235}$U, which minimizes the contributions of other nuclei
such as $^{239}$Pu and $^{238}$U to the fission rates. 
On the one hand, this could be an advantage for fundamental physics purposes as the antineutrino energy 
spectrum associated to the fission of $^{235}$U is the most precisely known. 
It also means that if neutrino detectors with a good energy resolution are used, experiments located close to research reactors could provide new measurements 
of the $^{235}$U antineutrino energy spectrum. On the other hand, it shows that the measurement of the contribution of $^{239}$Pu to the fissions with an antineutrino detector for non-proliferation purposes is  challenging in such a research reactor.

\begin{figure*}[hbt!]
\begin{center}
\includegraphics[scale=0.4]{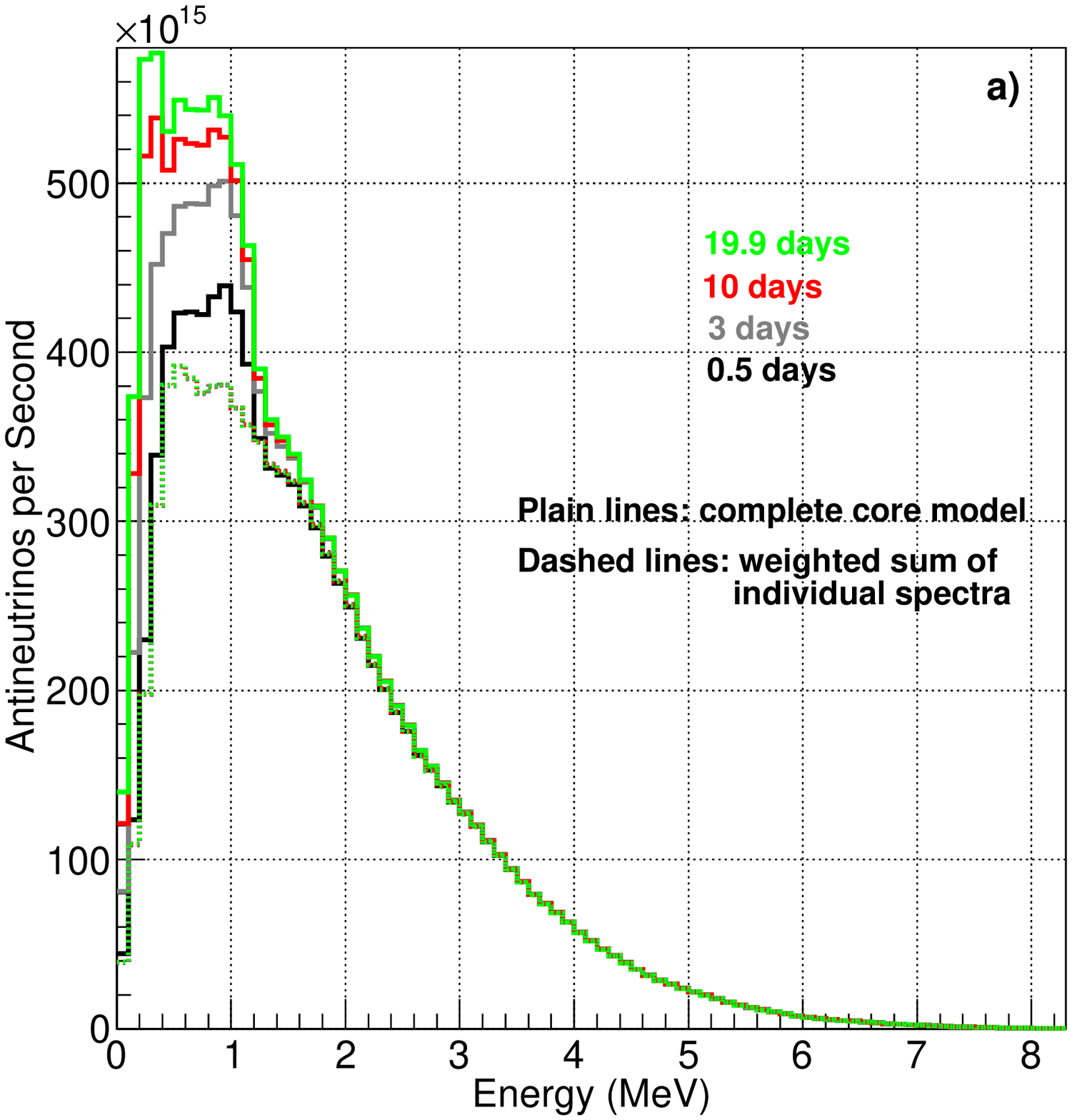}
\includegraphics[scale=0.4]{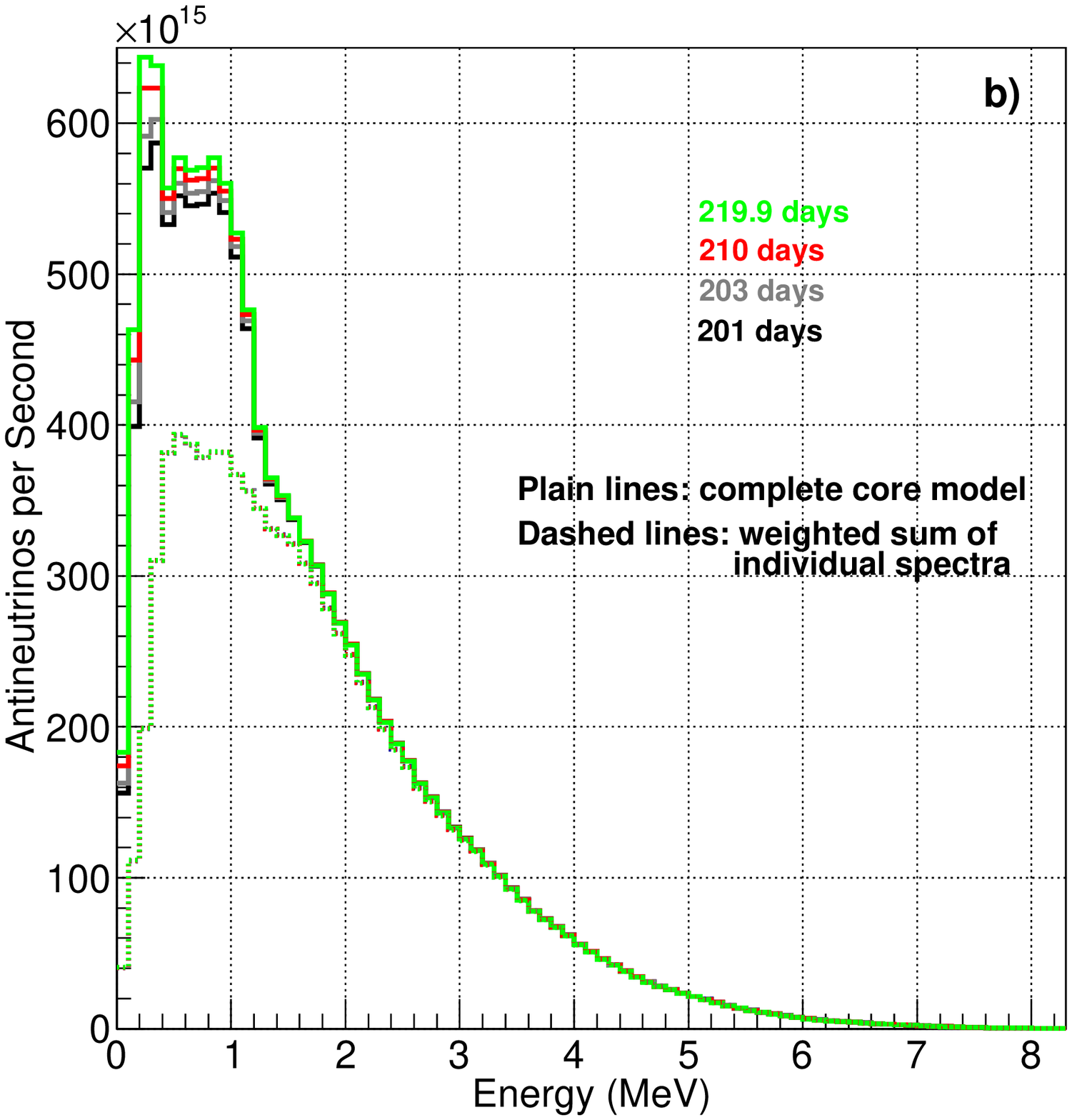}
\caption{\label{fig:EmittedEnergySpectra} 
Emitted antineutrino energy spectra computed with the complete core model (plain lines) and with individual summation method spectra for $^{235}$U, $^{239}$Pu, $^{238}$U and $^{241}$Pu weighted 
by the fission rates computed with the same core simulation (dashed lines) for the following time after core start-up: 
a) 0.5, 3\,days, 10\,days and 19.9\,days ; b) 201\,days, 203, 210 and 219.9\,days. Results for dashed lines are superposed.}
\end{center}
\end{figure*}

The emitted flux computed with the complete core model presented here connected with the summation method is much larger than when using the individual energy spectra of $^{235}$U, $^{239}$Pu, $^{238}$U and $^{241}$Pu, computed with the summation method after respectively 12, 36, 36 and 36\,hours of irradiation, weigthed with the fission rates arising from the same core simulation. This is due to the contribution of $^{239}$U and $^{239}$Np beta decays at low energy, as can be seen on the emitted energy spectra computed with both methods and presented in Fig.\,\ref{fig:EmittedEnergySpectra}\,a) and\,b).
 These two latter figures correspond to two sets of time steps of the core simulation. Fig.\,\ref{fig:EmittedEnergySpectra}\,a) displays the evolution of the antineutrino energy spectrum at the very beginning after the start-up of a core loaded with fresh fuel. 
The increase of the antineutrino energy spectra can be seen at 0.5, 3, 10 and 19.9\,days after start-up. The spectra displayed in plain lines show the summation calculation directly issued from the complete core simulation, while the spectra in dashed lines show the calculation using the individual spectra weighted by the fission rates from the same simulation. 
The power of the reactor core is kept constant in the simulation. One can notice that regarding the spectra computed using the individual uranium and plutonium spectra, 
the time evolution has no impact; the four dashed line spectra are superposed. This is awaited as these individual spectra have been computed without off-equilibrium effects or other reactor physics effects, and after a fixed irradiation time. The discrepancy between both calculations in terms of normalization and evolution with time is clearly evidenced, due to the effects mentioned above: contribution of actinides that can be accounted for only in a complete core simulation, and off-equilibrium effects which include the effect of neutron capture on the fission product concentrations and the time evolution of their concentrations in-core with respect to the chosen irradiation time of the individual spectra. These contributions may be different at reactors using a different type of fuel, exhibiting another neutron energy spectrum, or based on other designs such as power stations. Indeed the contributions of $^{239}$U and $^{239}$Np arise from successive neutron captures and beta decays starting from $^{238}$U at the origin of the creation of $^{239}$Pu. 
These contributions will be more pronounced in PWRs in which the creation of $^{239}$Pu is such that  
this isotope exhausts nearly half of the total fission rate at the end of a cycle. 
The discrepancy between emitted antineutrino spectra from both methods is even more pronounced for spectra computed for the core
 having reached its composition at equilibrium, as illustrated in Fig. \ref{fig:EmittedEnergySpectra}\,b) 
 which is the same as Fig.\,\ref{fig:EmittedEnergySpectra}\,a) but computed after 201, 203, 210 and 219.9\,days of 
 core evolution with a periodic refueling by seventh of the core every 20\,days. Here again, the spectra built with the individual spectra are superposed. 
 The spectra built with the complete core model are much more alike than before, showing that the evolution of 
 the fuel composition is weaker when the reactor has reached equilibrium.

In Appendix A, a table is provided containing the emitted antineutrino energy spectrum at 210 days 
after core start-up, computed with the sum of the contribution of all $\beta$-emitters in the complete core model, in order to allow the reader to use it as an input of the simulation of any specific detection setup. 
This could be useful for experiments chasing sterile neutrinos at research reactors. 
The energy spectrum can also be used in neutrino nucleus coherent scattering experiments, 
or any other experiment using another interaction process than the inverse beta decay\,\cite{NNCS, NeutrinoMagneticMoment}.

\subsection{Example of Antineutrino Energy Spectrum and Flux Detected by the IBD process}

\begin{figure}[hbt!]
\includegraphics[scale=0.4]{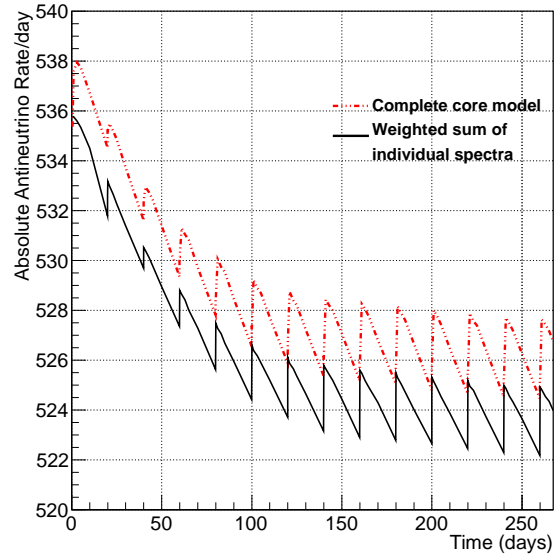}
\caption{\label{fig:DetectedAntinus} 
Antineutrino flux which would be measured in a detector with a 50\% efficiency placed at 7\,m from a research reactor 
with a $^{235}$U enriched fuel of 19.75\%. The red dashed and black plain lines correspond respectively to the complete core model method and 
to the individual spectra method.}
\end{figure} 

In Fig.~\ref{fig:DetectedAntinus} are displayed 
the antineutrino fluxes, computed with the two methods exposed above (in dashed line with the complete core model, in plain line with individual spectra), which would be detected in a detector filled in with 0.85~t of liquid scintillator doped with Gd 
at 7~m from the core and with a detection efficiency of 50\%. The energy spectra corresponding to Fig.\,\ref{fig:EmittedEnergySpectra}\,b) have been folded 
by the IBD process cross-section\,\cite{Vogel}, weighted by the afore mentioned solid angle and detection efficiency. 
No energy resolution nor energy dependence of the detection efficiency have been taken into account here, in order to show an example of result rather independent from the used experimental setup. 
The obtained variation of the detected antineutrino flux along a fuel cycle is only of 0.7\%. 
Though very small, this variation is not null, and a misuse of the reactor which would imply a larger $^{239}$Pu content in-core could result
 in a visible variation with a high efficiency detector, as shown with a toy model in ref.\,\cite{NuciferPaper}. 
 More detailed studies using our reactor model are contemplated, and beyond the scope of this article. 
%
A small discrepancy between the detected fluxes computed with the two methods presented above can be seen, which remains well below one percent. 
 Indeed, after folding with the IBD cross-section, the discrepancies nearly totally vanish and the two fluxes are 
 very similar.
  This is well understandable as the effects of the beta decays of actinides, neutron capture and evolution of the fission product concentrations are very much suppressed by the detection threshold of the IBD process. 
\begin{figure*}[hbt!]
\includegraphics[scale=0.4]{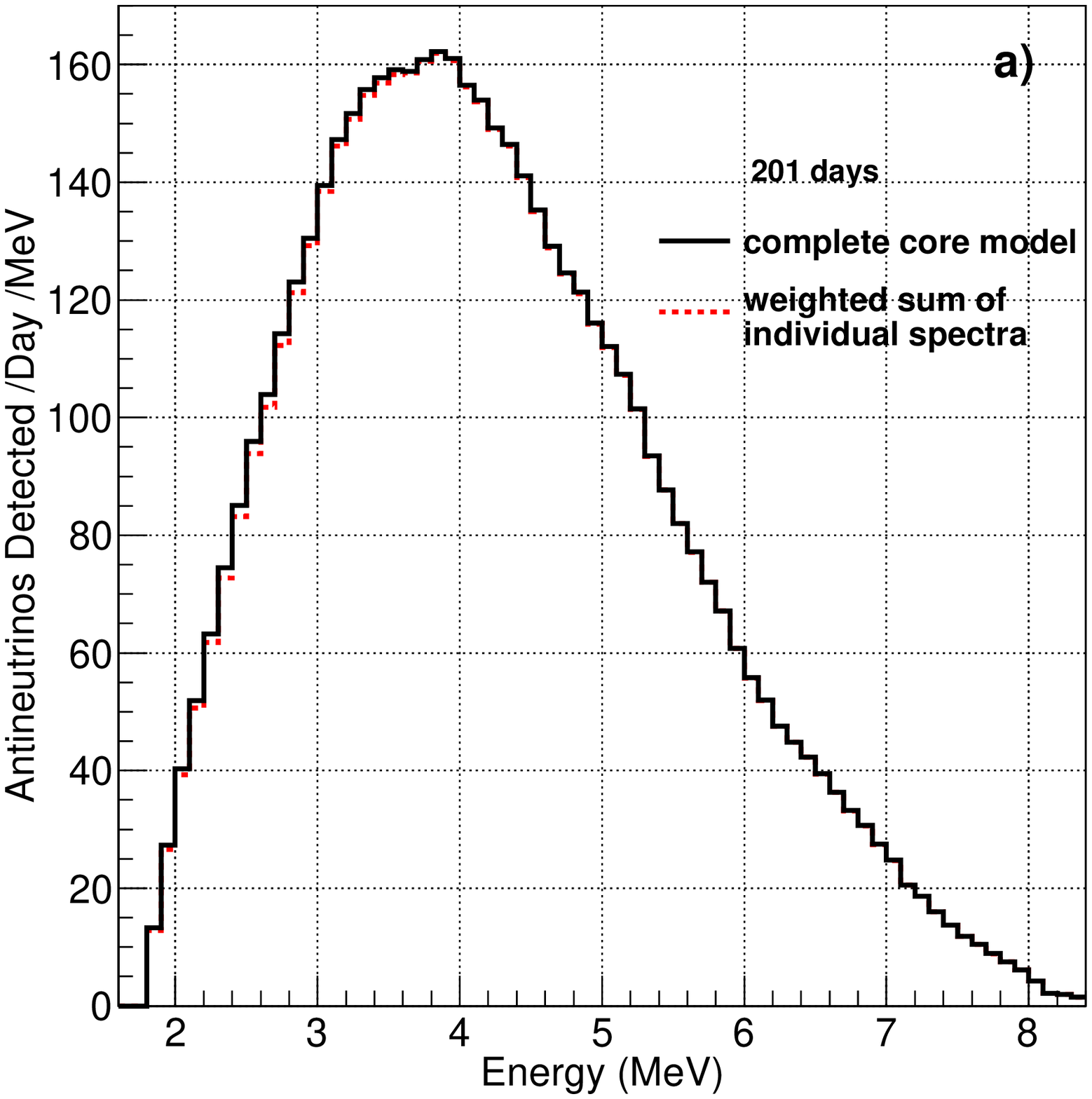}
\includegraphics[scale=0.41]{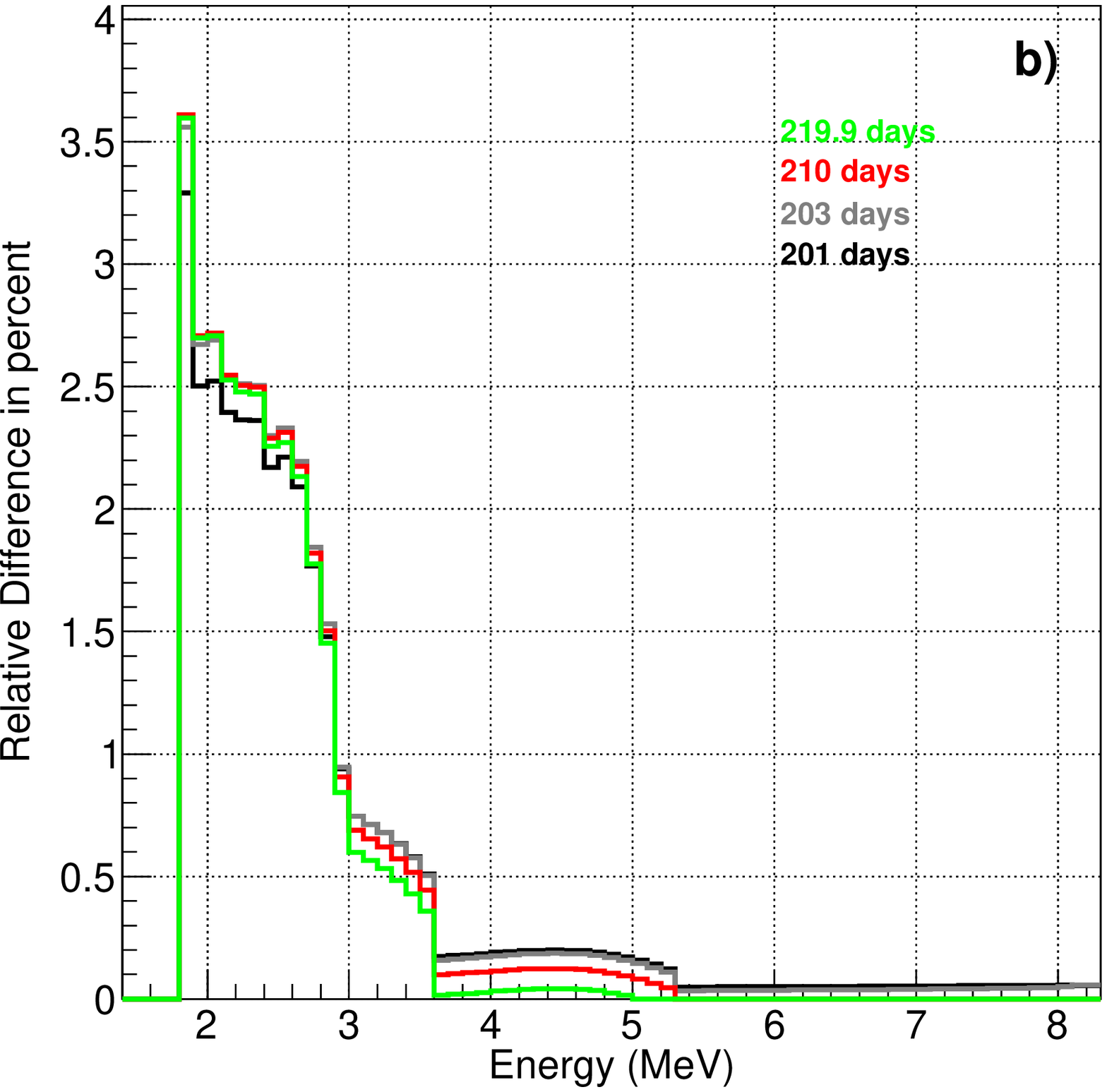}
\caption{\label{fig:DetectedEnergySpectra} 
a) The two spectra presented are the ones from Fig. \ref{fig:EmittedEnergySpectra}\,b) taken after 201 days of core operation and folded with the IBD cross-section, and detected in a detector placed at 7\,m from the reactor, 
with a target volume of 0.85\,m$^3$ of Gd-doped scintillator and assuming a 50\% detection efficiency. 
The plain line represents the spectrum computed with the complete reactor model coupled with the summation method, while the dashed line represents the spectrum built with the individual uranium and plutonium antineutrino spectra weighted by their contributions to the fissions in-core. b) Relative difference (spectrum from the complete core coupled to summation method - individual summation method spectra weighted by fission rates) over the spectrum from the complete core coupled to summation method, after folding by the IBD cross-section and in the same detection conditions as a).}
\end{figure*}

The detected antineutrino energy spectra in the conditions exposed above are displayed in Fig.\,\ref{fig:DetectedEnergySpectra} a), for a reactor considered at equilibrium (201\,days after core start-up), just after refueling, thus at the beginning of a new cycle. The black plain line represents the spectrum built with the distribution of the fission products from the OSIRIS core model, while the red dashed line represents the spectrum built using the individual spectra from the four main uranium and plutonium isotopes. The remaining discrepancy between the two spectra is located between 1.8 and 3.5\,MeV. Fig.\,\ref{fig:DetectedEnergySpectra} b) allows to highlight the relative difference among them, giving the contribution of 
actinide beta decay, neutron capture and fission product evolution after folding the spectra with the inverse beta decay cross-section, 
after 201\,days (in black), 203\,days (in grey), 210\,days (in red) and 219.9\,days (in green) of core evolution. One can see that the order of magnitude of these corrections
 with respect to individual spectra taken at the irradiation time of the ILL measurements by\,\cite{Schreckenbach} reaches 3.5\% at the detection threshold and lies between 2.0 and 2.4\% at 2.5\,MeV depending on the core irradiation time. These effects fall below 1\% after 3\,MeV.
The impact of the off-equilibrium effects is not negligible at the IBD detection threshold and is influenced by the short duration of the research reactor cycle (20 days in the case of OSIRIS) which imposes the renewal of one seventh of the fuel. 
Antineutrino experiments installed at research reactors using Low Enriched Uranium fuel will have to cope with such effects.
 Research reactors using High Enriched Uranium fuel should exhibit similar effects, except at the IBD threshold, where the beta decay of actinides have an impact. 
 Indeed, the contribution of the actinide beta decays should be much smaller due to the smaller quantity of $^{238}$U in-core.
In Appendix A, a table is provided containing the relative difference displayed in Fig.\,\ref{fig:DetectedEnergySpectra} b) after 210\,days of core operation.

\section{Gamma and Neutron background}
  In a reactor environment, the gamma sources are numerous: prompt gamma rays emitted when the fission process occurs, 
$\gamma$ decays of the fission fragments after their $\beta$$^{-}$ decays, (n,$\gamma$) reactions in the fuel or in different reactor components, and more generally gamma rays emitted by the excited nuclei 
created in the different processes. Using the attenuation coefficients available in the literature for water, air and concrete, the total attenuation of the gamma rays emitted and transported
 up to the Nucifer detector casemate was estimated to be about 10$^{8}$ at 8 MeV, 10$^{10}$ at 4 MeV and up to 10$^{21}$ at 1 MeV.
In order to study in details the production process of gamma rays and their propagation to the detector casemate with enough statistics,
the simulated pool and concrete walls were discretized in cells of 50~cm. A variance reduction technique was also used by weighting the importance of the different cells by the number of
$\gamma$ tracks. 
The chosen strategy was to perform simulations with the reactor core only (C), core and pool only (CP) and core-pool plus concrete walls (CPW) with the MCNPX 2.7.d
 code \cite{MCNPX} 
using preferentially the ENDFB/VI cross-section library and the ENDFB/VII one for photonuclear reactions plus a thermal S($\alpha$,$\beta$) treatment of the scattering reactions in the water \cite{ENDFBVI,ENDFBVII}. 
The purpose of these three different simulations was to understand the origin of the gamma rays and neutrons reaching the experiment location. 
This method also allowed to decrease the CPU cost, as the transportation of the gamma rays from the core through the pool and the concrete walls is challenging. 
%
The core model adopted in this study is the first one presented in section III: a core loaded with fresh fuel and all the fuel assemblies grouped in a unique universe.
As a first step in each of the three cases (C, CP, CPW), a simulation of reference using the default physics option of MCNPX 2.7.d for the $\gamma$ transport
 was performed in order to carry out the simplest simulation with the following reactions :
Compton scattering, pair production, Thomson and Rayleigh scatterings, photoelectric effect plus K and L shell flurorescences. 
In a second step, extra simulations were performed by adding only one additional physics phenomenon at a time: delayed gamma rays, Bremsstrahlung, photonuclear reactions, 
in order to estimate their respective contributions in the neutron and gamma production processes.
In order to propagate the source correctly in the geometry, a source is generated with a simulation using 50 000 source particles, 1 000 inactive cycles and 700 active cycles. The generated source is then used in 
the subsequent simulations, with 50 000 source particles, 20 inactive cycles and 470 active cycles. 

\begin{figure*}[hbt!]
\begin{center}
\includegraphics[scale=0.85]{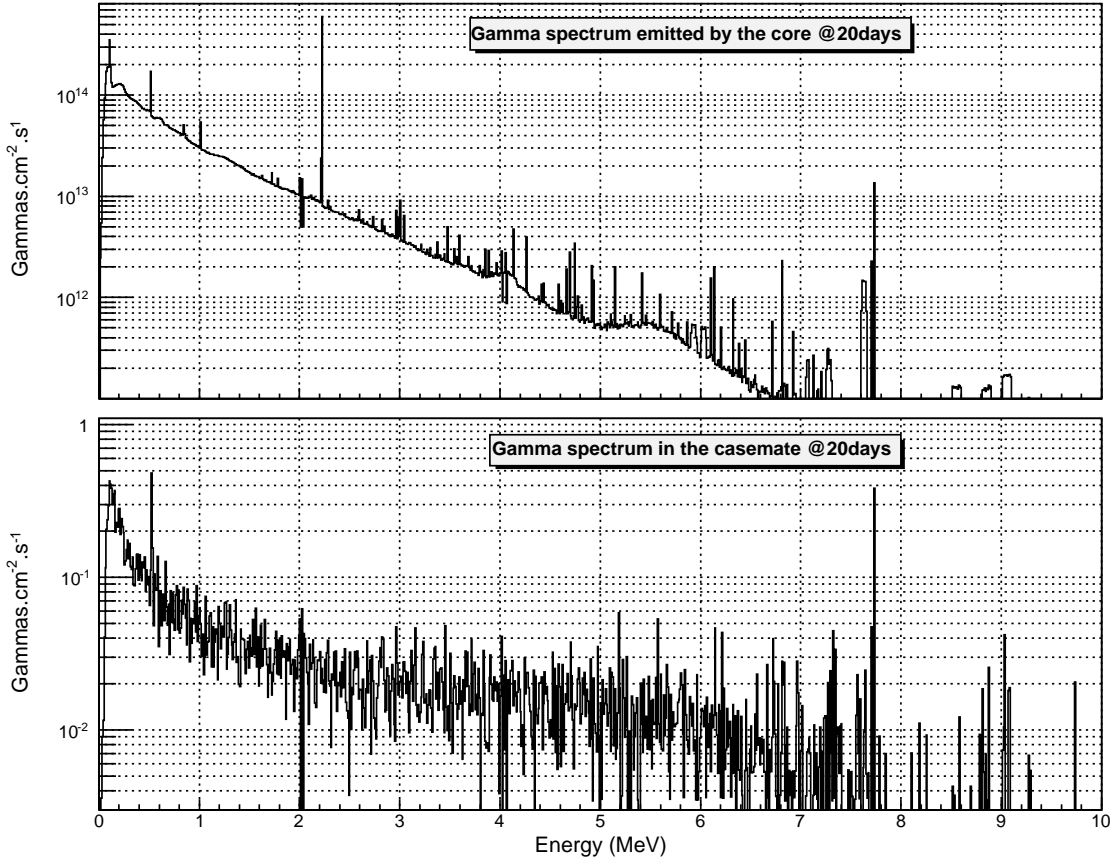}
\caption{\label{fig:GammaSpectrum} 
Gamma  spectra computed with MURE for a reactor fuel composition at the equilibrium after 20 days in the core (upper panel) 
and in the casemate (lower panel).}
\end{center}
\end{figure*} 
The final gamma spectrum calculated with MURE for a reactor fuel composition at equilibrium 
after 20 days and taking into account all these reactions is presented in Fig.~\ref{fig:GammaSpectrum}.
The upper part of the figure refers to the gamma rays produced in the reactor core, also tabulated in Appendix B, and the lower part to the gamma background expected in the detector casemate.
The size of the energy bins is 10 keV.
The electron transport is not directly included in these simulations. 
Instead, a Thick Target Bremsstrahlung model is used which assumes that the electrons are generated in the direction of the incident gamma ray
and immediately annihilated after generating Bremsstrahlung gammas. 
The 2.2 MeV gamma ray, attenuated in the concrete walls, corresponds to the (n,$\gamma$) radiative capture by hydrogen nuclei contained in water.
One of the most important gamma ray at 7.7 MeV is due to the radiative capture on $^{27}$Al which is included in the fuel cladding. 
Radiative captures on $^{9}$Be included in the reflector walls were also identified.
The neutron and photon transports performed with MCNPX 2.7.d after one time evolution step takes 24 hours 
using a 3 GHz Intel\textregistered~Xeon\textregistered~E5450 processor. 
%
The contribution of delayed gammas could also be included in our MCNPX 2.7.d calculation using either multigroup data either a multigroup mode coupled to energy lines 
if individual data are available \cite{MCNPX-GAMMA}.
The computing time needed for a multigroup mode coupled to energy line calculation added to the computing time needed for the fuel evolution
prevents such a detailed calculation.
However, the multigroup data calculation gives the general shape of the gamma energy spectrum and meets our needs with respect to the energy resolution of the NUCIFER detector.
In this case, the number of $\gamma$ tracks in the CPW simulation is 
multiplied by 5 and the computing time increased by 50\%.
%
K and L shell flurorescences contribute to 40\% of the total number of $\gamma$ tracks in the complete simulation. 
Most of the fluorescence reactions (85\%) occurs in the core where are located the elements with a 
high atomic number such as the fuel elements.
The respective contributions of the gamma rays created by Bremsstrahlung in the core, pool and casemate are
71\%, 27\% and 2\% with typically energies lower that 3 MeV and then could be neglected for the background computation in the detector casemate.

The neutrons emitted by the core are thermalized in the pool. 
The influence of the photonuclear reactions and especially the neutron production through the ($\gamma$,n) reactions is also studied.
Only photonuclear reactions with a reaction threshold below 8 MeV are considered because it corresponds to an important decrease in the gamma flux, 
as shown in Fig.~\ref{fig:GammaSpectrum}.
The comparison of C, CP and CPW simulations shows that 85\% of the photonuclear reactions occur in the concrete walls, 
mostly ($\gamma$,n) reactions on $^{40}$K. 
This secondary neutron production could partly explain the neutron background measured in the detector casemate by the Nucifer collaboration\,\cite{NuciferPaper} and which increases with the reactor power. 
Further studies are required to know the exact percentage of $^{40}$K in the concrete walls and the access to a detailed composition of the different concretes used in the OSIRIS facility, and are beyond the scope of this paper.

\begin{figure*}[hbt!]
\begin{center}
\includegraphics[scale=0.6]{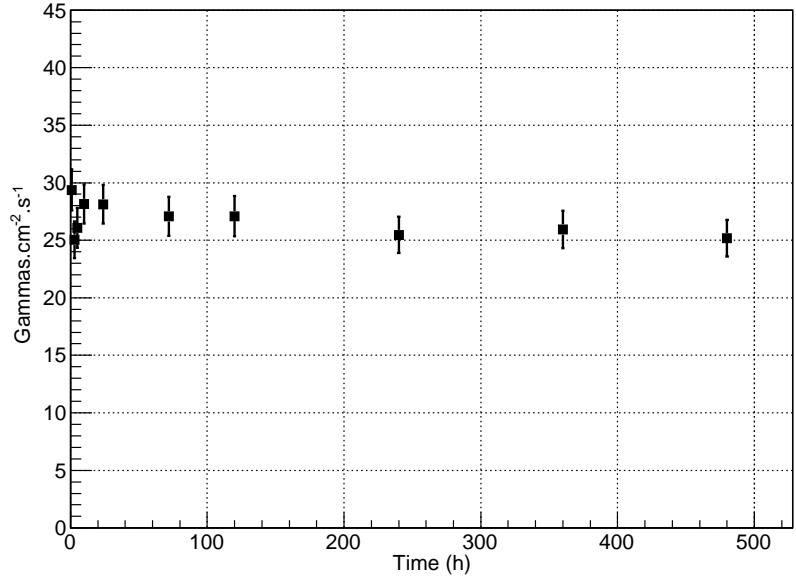}
\caption{\label{fig:GammaRate} Gamma flux rate in the detector casemate as a function of time, computed with MURE from a reactor fuel 
composition at the equilibrium.}
\end{center}
\end{figure*}

In Fig.~\ref{fig:GammaRate}, the gamma flux rate expected in the detector casemate is calculated with MURE for a reactor cycle of 20 days 
considering a fuel composition at equilibrium obtained with the refueling scenario presented in the section III.
No energy threshold is applied on the gamma detection. 
The error bars are statistical and computed using the number of $\gamma$ tracks.
From this calculation, the gamma flux rate in an antineutrinos detector of 10 m$^{2}$ surface is estimated to be 2.5 MHz.
A more realistic fuel composition could be used in the future to improve the normalisation of the gamma flux rate but this first result is already in good
agreement with the order of magnitude of 5 MHz measured in the casemate by the NUCIFER collaboration\,\cite{NuciferPrivate}.

\section{Conclusions}
%
The simulation of the OSIRIS research reactor (CEA-Saclay, France),  was performed with the MURE code, including the core, the pool and a detector casemate located behind concrete walls where the Nucifer antineutrino detector has been placed at 7\,m from the core.
The implementation of the periodic refueling of the core every 20 days by seven-th made it possible to carry out a first study of the emitted antineutrino spectrum which could be measured at such a research reactor.
This core simulation was benchmarked with the neutron energy spectrum calculated in an irradiation cell 
by the deterministic code APOLLO2. 
The antineutrino energy spectrum and flux were computed using the summation method, coupled for the first time to the modelization of a reactor core evolution. The variation of the detected antineutrinos along a 20 days operation cycle was calculated.
It was shown that the contribution of $^{239}$Pu and $^{241}$Pu to the antineutrinos spectrum is weak, due to the 20\% fuel enrichment in $^{235}$U, 
which is a common enrichment at research reactors. The emitted antineutrino energy spectrum associated to the reactor fuel at equilibrium is provided and could be used by experiments studying fundamental properties of neutrinos at such facilities.
Off-equilibrium effects affecting the antineutrino spectral shape due to neutron capture and time evolution of the fission product concentrations in-core were computed, as well as the impact of actinides created along the operation of the reactor. 
The obtained corrections could be applied to antineutrino energy spectra computed by conversion of the ILL reference beta spectra\,\cite{SchreckU5-1,SchreckU5-2,SchreckU5Pu9,Schreckenbach} 
like the ones from\,\cite{Huber} 
in the frame of any sterile neutrino experiment located at a research reactor using a similar fuel enrichment.
In a second part, a detailed study of the gamma ray background was performed.
This work allowed a better understanding of: the contribution of the fluorescence reactions of fuel, 
the importance of Bremsstrahlung reactions for the gamma rays present in core and their negligible contribution in the casemate, 
and the origin of the neutron background measured in the casemate. The latter does not come directly from the core but rather from ($\gamma$,n)
reactions on $^{40}$K nuclei present in the concrete walls. Overall the model succeeded to reproduce the gamma background rate level measured at the location of the Nucifer detector.  
The gamma energy spectra produced by the reactor core, and after propagation up to the detector location are provided, as these results could be used to compute the sensitivity of short baseline reactor neutrino experiments 
foreseen in the frame of the search of potential sterile neutrinos.

\section{Acknowledgements}

The authors thank the NEEDS challenge and the GEDEPEON research groupment 
for their financial support. The authors thank the people from the OSIRIS reactor for providing useful information.



\begin{widetext}
\appendix  

\section{ANTI-NEUTRINO EMISSION FROM THE OSIRIS RESEARCH REACTOR} 
\subsection{Anti-neutrino rate emitted by the OSIRIS research reactor, computed with the complete core model with the MURE code and the summation method, for a reactor fuel 
composition at equilibrium, after 210 days of irradiation and a periodic refueling every 20 days. The total fission rate in-core at this irradiation time is 2.16007\,$10^{18}$ fissions per second.}
 \centering

\subsection{Relative difference between the anti-neutrino energy spectrum computed with the full reactor simulation and the one computed with individual uranium and plutonium spectra weigthed by their fission rate, for a fuel at equilibrium after 210 days of operation after core start-up.}

 \centering

  \section{GAMMA SPECTRUM IN THE CORE, COMPUTED WITH MURE FOR A REACTOR FUEL COMPOSITION AT THE EQUILIBRIUM AFTER 20 DAYS.} 
  
 \centering
\end{widetext}


\begin{thebibliography}{9}

\bibitem{refsNeutrinoProp} J. Beringer {\it et al.}, Particle Data Group, Phys. Rev.  {\bf D86}, 010001 (2012).
\bibitem{refNeutrinoMoni} V. A. Korovkin {\it et al.}, Atom. Ener. {\bf 65}, 3, 712 (1988).

\bibitem{IAEAReport} IAEA Report SG-EQGNRL-RP-0002 (2012).
\bibitem{PRLHuber} E. Christensen, P. Huber, P. Jaffke and T. Shea, Phys. Rev. Lett. {\bf 113}, 042503 (2014).
\bibitem{Mueller} Th. Mueller  {\it et al.}, Phys. Rev.  {\bf C83}, 054615 (2011).
\bibitem{Huber}   P. Huber, Phys. Rev.  {\bf C84}, 024617 (2011).
\bibitem{Mention} G. Mention {\it et al.}, Phys. Rev.  {\bf D83}, 073006 (2011).
\bibitem{WhitePaper} K. N. Abazajian {\it  et al.}, http://arxiv.org/abs/1204.5379.
\bibitem{RevueNonProlif} M. Fallot, Nuclear Data Sheets {\bf 120}, 137(2014).
\bibitem{NNCS} A. Drukier and L. Stodolsky, Phys. Rev. {\bf D 30}, 2295 (1984).
\bibitem{NeutrinoMagneticMoment} V. I. Kopeikin, L. A. Mikaélyan, V. V. Sinev, Physics of Atomic Nuclei
 {\bf 63}, Issue 6, 1012 (2000).
\bibitem{NuciferPaper} {Online Monitoring of the Osiris Research Reactor with the Nucifer Neutrino Detector, 
the NUCIFER collaboration, ArXiv 1509.05610, 2015}.
\bibitem{IAEA} \url{https://www.iaea.org/OurWork/ST/NE/NEFW/Technical-Areas/RRS/conversion.html}.

\bibitem{SchreckU5-1}  K.~Schreckenbach {\it et al.}, Phys. Lett. {\bf 99B}, 251 (1981).
\bibitem{SchreckU5-2}  K.~Schreckenbach {\it et al.}, Phys. Lett. {\bf 160B}, 325 (1985). 
\bibitem{SchreckU5Pu9} F.~von~Feilitzsch, A.~A.~Hahn and K. Schreckenbach, Phys.\ Lett.\ {\bf 118B}, 162 (1982).
\bibitem{Schreckenbach} A.A. Hahn {\it et al.},  Phys. Lett. {\bf B218} (1989) 365.

\bibitem{PRLHaag} N. Haag {\it et al.}, Phys. Rev. Lett. {\bf 112}, 122501 (2014).







\bibitem{Hayes} A. C. Hayes {\it et al.}, Phys. Rev. Lett. {\bf 112}, 202501 (2014).

\bibitem{Neutrino2014} Double Chooz and Reno presentations at the XXVI International conference on Neutrino Physics and Astrophysics, Boston (2014) conference \url{http://neutrino2014.bu.edu/}, Daya Bay presentation at ICHEP 2014 conference, \url{http://ichep2014.es/}.
 

\bibitem{DC} Y. Abe {\it et al.}, Phys. Rev. {\bf D 86}, 052008 (2012).

\bibitem{Reno} J.K. Ahn {\it et al.}, Phys. Rev. Lett. {\bf 108}, 191802 (2012). 

\bibitem{DB} F. P. An {\it  et al.}, Phys. Rev. Lett.  {\bf 108}, 171803 (2012).

\bibitem{Fallot}  M. Fallot  {\it et al.}, Phys. Rev. Lett  {\bf 109}, 202504 (2012).

\bibitem{PRL2015} A.-A. Zakari {\it et al.}, Phys. Rev. Lett. {\bf 115}, 102503 (2015).

\bibitem{OSIRIS} \url{http://www.cad.cea.fr/rjh/Add-On/osiris_gb.pdf}. 

\bibitem{MURE} MURE, MCNP utility for reactor evolution (2009), 
\url{http://www.nea.fr/tools/abstract/detail/nea-1845}. 
O. Meplan, Tech. Rep. LPSC  {\bf 0912} and IPNO-09-01 (2009).
\bibitem{MCNPX} \url{https://mcnpx.lanl.gov/}.

\bibitem{Malouch} F. Malouch, Th\`{e}se de Doctorat, Rapport CEA {\bf R-6034} (2003).
\bibitem{Kopeikin} V. Kopeikin {\it et al.}, Phys. At. Nucl. {\bf 67}, 1892 (2004).

\bibitem{ENDFBVI} V. McLane {\it et al.}, ENDF/B-VI Summary Documentation Supplement I, ENDF/HE-V1 Summary Documentation (1996).

\bibitem{Jones} C. Jones {\it et al.}, Phys. Rev. {\bf D86 }, 012001 (2012).
\bibitem{SimuDC} Fission rates calculation for the Double Chooz experiment, A. Onillon {\it et al.}, in preparation.
%


\bibitem{Apollo2} R. Sanchez {\it et al.}, Nucl. Sc. and Eng., Vol. {\bf 100}, 352 (1988). 
\bibitem{JEF2} OECD/NEA Data Bank, 2000. The JEF-2.2 Nuclear Data Library. JEFF Report 17. NEA/OECD Publications, Paris.

\bibitem{RefCommOSiris} Private communication, CEA/DEN/OSIRIS, 2011.

\bibitem{Algora} A. Algora {\it et al.}, Phys. Rev. Lett. {\bf 105}, 202501 (2010).



\bibitem{Vogel} P. Vogel and J. F. Beacom, Phys. Rev. {\bf D60}, 053003 (1999).
\bibitem{ENDFBVII} M. B. Chadwick {\it et al.}, Nucl. Data Sheets {\bf 107}, 2931 (2006).
\bibitem{MCNPX-GAMMA} J. W. Durkee Jr. {\it et al.}, Prog. in Nucl. En. {\bf 51}, 813 (2009).
\bibitem{NuciferPrivate} Private communication, the NUCIFER collaboration, 2013.


\end{thebibliography}
\end{document}